\documentclass[twocolumn]{aastex631}
\usepackage{CJK}

\usepackage{tabularx}
\usepackage{hyperref}
\hypersetup{
    colorlinks=true,
    allcolors=cyan
    }

\shorttitle{KLIP Optimization for MagAO GAPlanetS}
\shortauthors{Adams, Follette, Wang et al.}

\graphicspath{{./}{figures/}}

\begin{document}

\begin{CJK*}{UTF8}{gbsn}

\title{The Giant Accreting Protoplanet Survey (GAPlanetS): Optimization Techniques for Robust Detections of Protoplanets}

\correspondingauthor{J\'ea I. Adams Redai}
\email{jea.adams@cfa.harvard.edu}
\author[0000-0002-4489-3168]{J\'ea I. Adams Redai}
\affiliation{Center for Astrophysics $|$ Harvard \& Smithsonian, 60 Garden St, Cambridge, MA 02138, USA}
\affiliation{Amherst College Department of Physics and Astronomy, PO Box 5000, Amherst, MA 01002-5000, USA}

\author[0000-0002-7821-0695]{Katherine B. Follette}
\affiliation{Amherst College Department of Physics and Astronomy, PO Box 5000, Amherst, MA 01002-5000, USA}

\author[0000-0003-0774-6502]{Jason Wang (王劲飞)}
\altaffiliation{51 Pegasi b Fellow}
\affiliation{Department of Astronomy, California Institute of Technology, Pasadena, CA 91125, USA}

\author{Clare Leonard}
\affiliation{Amherst College Department of Physics and Astronomy, PO Box 5000, Amherst, MA 01002-5000, USA}
\affiliation{Epic Systems, 1979 Milky Way, Verona, WI 53593 }


\author[0000-0001-6396-8439]{William Balmer}
\affiliation{Department of Physics \& Astronomy, Johns Hopkins University, 3400 N. Charles Street, Baltimore, MD 21218, USA}
\affiliation{Space Telescope Science Institute, 3700 San Martin Drive, Baltimore MD 21218, USA}
\affiliation{Amherst College Department of Physics and Astronomy, PO Box 5000, Amherst, MA 01002-5000, USA}

\author[0000-0002-1384-0063]{Laird M. Close}
\affiliation{Steward Observatory, University of Arizona, Tucson AZ 85721, USA}

\author{Beck Dacus}
\affiliation{Amherst College Department of Physics and Astronomy, PO Box 5000, Amherst, MA 01002-5000, USA}

\author{Jared R. Males}
\affiliation{Steward Observatory, University of Arizona, Tucson AZ 85721, USA}

\author[0000-0002-1384-0063]{Katie M. Morzinski}
\affiliation{Steward Observatory, University of Arizona, Tucson, 933 N Cherry Ave, Tucson, AZ 85721, USA}

\author{Joseph Palmo}
\affiliation{Amherst College Department of Physics and Astronomy, PO Box 5000, Amherst, MA 01002-5000, USA}

\author[0000-0003-3818-408X]{Laurent Pueyo}
\affiliation{Department of Physics \& Astronomy, Johns Hopkins University, 3400 N. Charles Street, Baltimore, MD 21218, USA}
\affiliation{Space Telescope Science Institute, 3700 San Martin Drive, Baltimore MD 21218, USA}

\author{Elijah Spiro}
\affiliation{Amherst College Department of Physics and Astronomy, PO Box 5000, Amherst, MA 01002-5000, USA}
\affiliation{NASA Kennedy Space Center}

\author[0000-0003-0660-9776]{Helena Treiber}
\affiliation{Amherst College Department of Physics and Astronomy, PO Box 5000, Amherst, MA 01002-5000, USA}


\author[0000-0002-4479-8291]{Kimberly Ward-Duong}
\affiliation{Department of Astronomy, Smith College, Northampton MA 01063 USA}

\author{Alex Watson}
\affiliation{Amherst College Department of Physics and Astronomy, PO Box 5000, Amherst, MA 01002-5000, USA}

\begin{abstract}

High-contrast imaging has afforded astronomers the opportunity to study light directly emitted by adolescent (tens of Myr) and ``proto" ($<$10Myr) planets still undergoing formation. Direct detection of these planets is enabled by empirical Point Spread Function (PSF) modeling and removal algorithms. The computational intensity of such algorithms, and their multiplicity of tunable input parameters, has led to the prevalence of ad-hoc optimization approaches to high-contrast imaging results. In this work, we present a new, systematic approach to optimization vetted using data of the high-contrast stellar companion HD 142527 B from the Magellan Adaptive Optics (MagAO) Giant Accreting Protoplanet Survey (GAPlanetS). More specifically, we present a grid search technique designed to explore three influential parameters of the PSF-subtraction algorithm \texttt{pyKLIP}-- annuli, movement, and KL modes. {We consider multiple metrics for post-processed image quality in order to optimally recover at H$\alpha$ (656nm) synthetic planets injected into contemporaneous continuum (643nm) images.} These metrics include: peak (single-pixel) SNR, average (multi-pixel average) SNR, 5$\sigma$ contrast, and false-positive fraction. We apply continuum-optimized KLIP reduction parameters to six  H$\alpha$  direct detections of the low-mass stellar companion HD142527 B, and recover the companion at a range of separations. Relative to a single-informed, non-optimized set of KLIP parameters applied to all datasets uniformly, our multi-metric grid search optimization led to improvements in companion SNR of up to 1.2$\sigma$, with an average improvement of 0.6$\sigma$. Since many direct imaging detections lie close to the canonical 5$\sigma$ threshold, even such modest improvements may result in higher yields in future imaging surveys.

\end{abstract}


\section{Introduction} \label{sec:intro}

Over the past decade, high contrast direct imaging has uncovered dozens of {bound sub-stellar companions to higher mass stars} \citep{bowler2016, currie2022}. {This technique is generally sensitive to faint companions at separations of $>0\farcs1$, and masses on the order of several Jupiter masses or larger.} Imaging's ability to resolve the light emitted directly by exoplanet atmospheres makes it a powerful vehicle for planet characterization. Consequently, prospects for future work constraining planet composition, formation, and habitability are intertwined with refinement of the imaging techniques that will allow us to robustly isolate planetary signals \citep{seager2010, Biller_2018}. 

Both current and future imaging campaigns are dependent on suppression of the stellar point spread function (PSF). Raw data of imaged extrasolar systems are dominated by the diffraction limited core of the stellar PSF, its broader seeing halo, and a field of uncorrected stellar ``speckles". This leaves faint planets buried under the starlight in raw and conventionally-combined data. {The hardware (adaptive optics, coronagraphs, apodizers, etc.) in some high-contrast imaging instruments, such as the Gemini Planet Imager \citep[GPI,][]{Macintosh2014} on the Gemini South telescope  and the Spectro-Polarimetic High contrast imager for Exoplanets REsearch \citep[SPHERE,][]{Bezuit2019} on the Very Large Telescope (VLT),  
can suppress starlight and allow for raw planet/star contrasts (i.e. the 5$\sigma$ noise level at planetary separations) of $\sim10^{-4}-10^{-5}$  \citep{bailey2016}.} Yet, detection of  young planets in near-infrared thermal emission requires planet/star contrasts of at least $10^{-6}$. 

Post-processing techniques, specifically PSF-subtraction, can improve planet/star contrasts by a factor of 10 -- 100 (see e.g. \cite{bailey2016}). A range of algorithmic approaches are available to achieve this improved contrast, the most common of which are the Karhunen-Lo\'eve Image Processing \citep[KLIP,][]{soummer2012, pueyo2016} and Locally-Optimized Combinations of Images \citep[LOCI,][]{Lafreniere2007} techniques. Both techniques utilize Angular Differential Imaging (ADI) \citep{Marois2006}, in which the source is allowed to rotate in the image plane throughout the observation, while the instrumental PSF remains static. On-sky rotation of the source ensures that PSF features identified by the algorithms consist primarily of instrumental PSF features, and exclude rotating high spatial frequency sources such as planets and narrow disk structures. In this work, we focus strictly on the Principal Component Analysis (PCA) based technique, KLIP, implemented with the python package \texttt{pyKLIP}\footnote{\url{https://bitbucket.org/pyKLIP/pyklip}} \citep{wang2015}. Optimization of LOCI algorithms is discussed in \citet{Thompson2021}.


Despite their power to suppress the stellar PSF, post-processing algorithms like KLIP are complex and highly tunable. Extracted photometry, astrometry, and spectroscopy of an exoplanet direct detection is greatly influenced by user-selected input parameters. The widely-used KLIP algorithm \texttt{pyKLIP}, for example, utilizes 25 tunable input parameters that control features such as application of a highpass filter, the complexity of the PSF model, the number and shape of regions for which model PSFs are constructed separately, and the size of the library of reference images. 

Of particular concern regarding KLIP parameter choices are systems containing \textit{both} planets and circumstellar material. Substructure is ubiquitous in planet-forming disks \citep{Benisty2022} and disk features have the potential to appear planet-like in post-processed images \citep[e.g.][]{follette2017}. A number of reported (proto)planet detections in these systems have been called into question when other techniques or datasets fail to reveal unambiguous planetary signals. These include the protoplanet candidates LkCa15 b and c \citep{kraus_ireland_2012, Sallum2015}, which have been contested by several papers highlighting their proximity to inner disk material \citep{curry2019, thalman2016}. Similarly, the two planet candidates in the HD 100546 system \citep{quanz, curry2015} were flagged as potential false positives because they were not recovered as orbiting point sources in other observations \citep{rameau2017, follette2017}.

We hypothesize that one cause of detection discrepancies among reported planet candidates is a lack of standardization in selecting PSF-subtraction parameters. The \textit{de facto} technique among the imaging community has been to make default choices for algorithmic parameters and to hand-tune those parameters once an apparent detection is made, or to optimize select parameters individually \citep{Meshkat2014}. These approaches are used in part because of the computational intensity of the KLIP algorithm, which does not lend itself well to optimization approaches requiring thousands to millions of iterations. 

{In this work, we use H$\alpha$ direct imagery of the $12 - 23$ au separation $0.26^{+0.16}_{-0.14}~\mathrm{M_\odot}$ \citep{Claudi2019} companion HD142527 B to develop a \texttt{pyKLIP} optimization methodology. This well-characterized companion has been observed as part of  the Giant Accreting Protoplanet Survey \cite{folletteinprep} over a long time baseline (2013 - 2018), and appears at a wide range of planet-star separations, making it more difficult to recover in some epochs than others. }

{This paper is organized as follows. We describe \texttt{pyKLIP} parameters of interest in Section \ref{paramsofinterest}. In Section \ref{methods}, we detail our Magellan Adaptive Optics (MagAO) observations and basic data processing procedures. In Section \ref{image_processing}, we outline our optimization approach, develop image quality metrics, and describe our selection of some fixed \texttt{pyKLIP} parameters. We present the final results of our optimizations in Section \ref{results}. Finally, we summarize our process and describe future steps in Section \ref{conclusion}} 

\begin{figure*}
    \begin{center}
        \includegraphics[width=\textwidth, keepaspectratio]{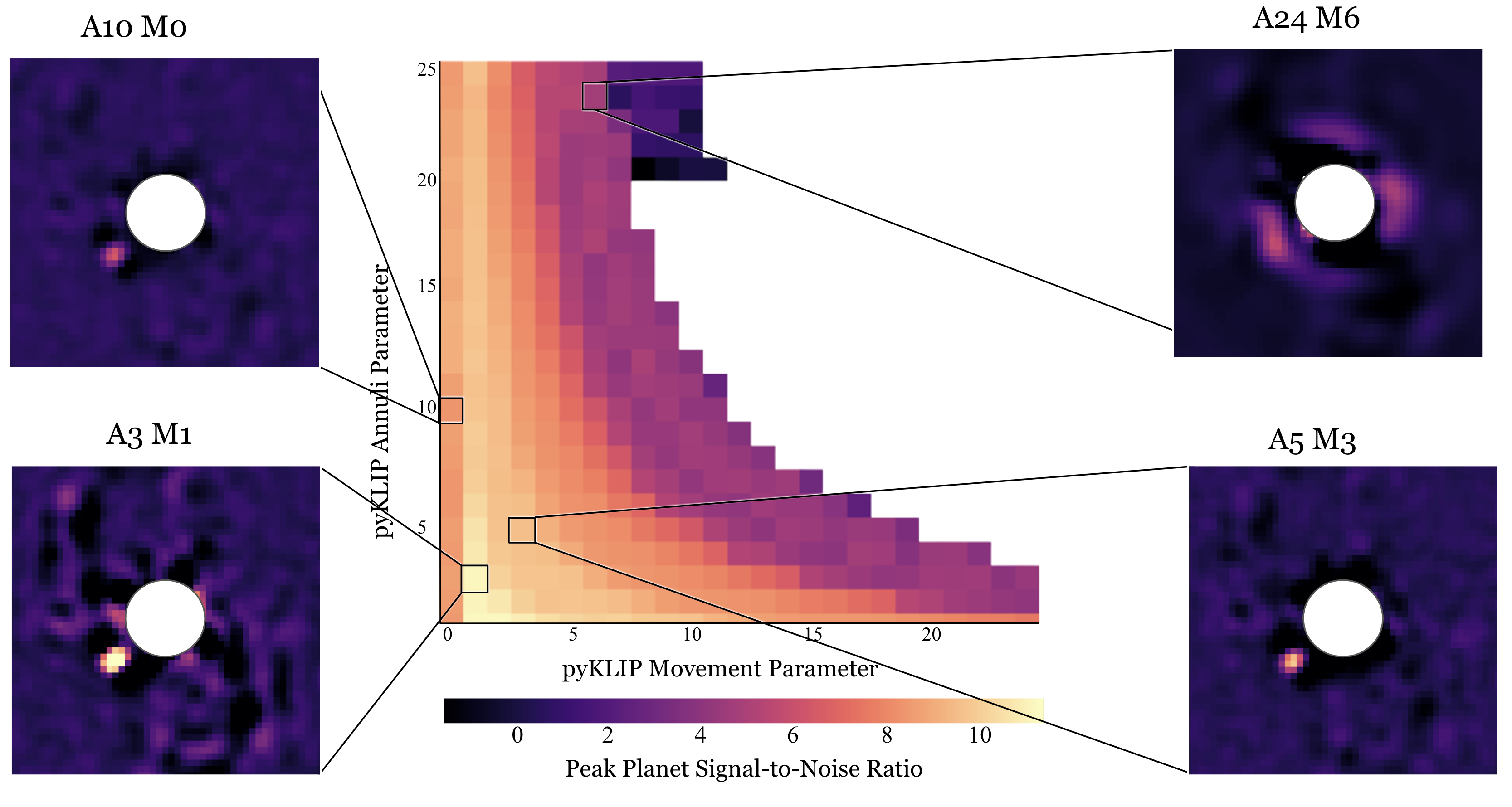}
    \caption{A depiction of \texttt{pyKLIP} output demonstrating the variation in the quality of recovery for a true companion (HD 142527~B) across a range of KLIP \texttt{movement} and \texttt{annuli} parameters. 
    Each pixel in the central heatmap represents a separate, independent KLIP reduction with the \texttt{movement} and \texttt{annuli} values depicted on the x and y axes, respectively. The color of that pixel reflects the highest single pixel SNR value at the location of the companion in the SNR map. Inset images depict the SNR maps for a representative sample of individual reductions. While much of this parameter space yields recovery of the companion at $>$5$\sigma$, there is intense variation in the quality of the extraction, and even some parameters for which the signal from the bright companion is nearly absent. For this dataset, an optimal combination of parameters to maximize the SNR of the companion for 10 KL modes is a \texttt{movement} value of 1~pixel and 3 \texttt{annuli}. {An in-depth explanation of this grid search technique is provided in Section \ref{pyKLIPPE}}. 
    }
    \label{fig:paramsmatter}
    \end{center}
\end{figure*}

\section{KLIP Parameters of Interest} \label{paramsofinterest}

All KLIP implementations rely on two fundamental elements of the algorithm: 1) The compilation of images to form a reference library, and 2) The construction of a custom PSF model for each image or region of an image from its PSF library, the complexity of which is controlled by the number of principal components (``KL modes"). ``KL modes" are a set of orthogonal basis vectors oriented to describe the variance in the reference library images, with each additional mode describing a smaller proportion of the overall variance. They can be thought of as common patterns in the images. As the number of KL modes used to build a PSF model increases, the patterns identified appear in fewer of the reference images. Therefore, higher KL modes correspond to a more ``aggressive" PSF subtraction. Of the 25 tunable user input parameters to the \texttt{pyKLIP} algorithm, those that have a particularly marked effect on the quality of post-processed images (even for bright companions, as shown in Figure \ref{fig:paramsmatter}) are:

\begin{enumerate} 
    \item The {\texttt{numbasis}} parameter controls the number of ``KL modes" that form the PSF model for subtraction, as described above. This parameter can be a single value or a list of values, in which case multiple PSF models will be created and subtracted to form multiple post-processed images. The maximum number of KL modes that can be used to construct a PSF for a given image sequence is equal to the number of images in the reference image set (for ADI imagery, this is the number of images in the sequence minus any that were discarded through rotational masking (see below)).  
    
    \item The {\texttt{annuli}} parameter controls the radial geometry of the optimization regions, specifically the number of annular zones within the image for which PSFs are constructed separately. In general, optimizing on smaller regions (a higher \texttt{annuli} parameter) means optimizing on fewer PSF features at once. These features include quasi-static speckles, the adaptive optics control radius, wind residuals, etc.
     
    \item The {\texttt{movement}} parameter is effectively an ADI rotational mask. It sets the number of pixels that an astrophysical source (e.g., planet) located at the center of each annular zone is required to have rotated relative to the image for which the PSF is being constructed (the ``target image") in order for another image in the sequence (a ``reference image") to be included in the reference library. A smaller \texttt{movement} value corresponds to more ``aggressive" subtraction, since more reference images with the planet located near the same position as in the target image are included in the reference library.

    \item The \texttt{highpass} parameter improves starlight subtraction by attenuating low spatial frequency signals in an image before executing KLIP.  \texttt{pyKLIP}'s Gaussian highpass filter parameter controls the standard deviation of this filter, where a small standard deviation is considered aggressive. Very little low spatial frequency signal will survive an aggressive highpass filter, however planet light may also be attenuated in this process.  
    
\end{enumerate}

\section{Data and Pre-processing} \label{methods}

\begin{deluxetable*}{ccccccccc}
    \tablehead{
    \colhead{{Date}} &  \colhead{{Sep. }} &  \colhead{{PA }}  & \colhead{{Seeing}} &\colhead{{$\mathrm{ t_{int}}$}}
    & \colhead{{$\mathrm{r_{sat}}$}} & \colhead{{Rot.}} & \colhead{{$\mathrm{ N_{ims}}$}}  & \colhead{{FWHM}}  \\   
    \colhead{} & \colhead{{(mas)}} & \colhead{{(deg)}} & \colhead{} & \colhead{{(min)}} & \colhead{{(pix)}} & \colhead{{deg}} & \colhead{} & \colhead{{(pix)}}
    }
    \startdata
    11 April 13  & 81.08 $\pm$ 1.08   & 128.12 $\pm$ 0.49 & 0.56  & 74.2 & 6 & 65.3 & 1961 & 4.56\\
    8 April 14  & 77.70 $\pm$ 1.68 & 117.01 $\pm$ 1.12 & N/A & 66.5 & N/A & 101.7 & 1758 & 4.00\\
    15 May 15  & 70.16 $\pm$ 1.19 & 110.56 $\pm$ 0.80 &  0.55 & 90.3 & N/A & 117.4 & 2387 & 5.50\\
    16 May 15  & 72.19 $\pm$ 2.02 & 107.84 $\pm$ 0.97 &  0.80 & 43.2 & 2 & 34.8 & 1143 & 5.01\\
    18 May 15  & 70.00 $\pm$ 1.35 & 110.12 $\pm$ 0.72 &  0.66 & 79.5 & 9 & 76.8 & 159 & 5.24\\
    27 April 18  & 44.34 $\pm$ 1.81 & 58.62 $\pm$ 1.67 &  N/A & 48.3 & 3 & 49.2 & 580 & 4.37\\
    \enddata
    \label{Tab:obs}
    \caption{Table of GAPlanetS observations of HD142527 B used in this analysis. Orbital locations were derived from \citep{Balmer2022}}.
\end{deluxetable*}

\subsection{Observational Data} \label{obs}

The data used in these analyses were taken with the Magellan Clay Telescope at Las Campanas Observatory using the Magellan Adaptive Optics \citep[MagAO,][]{Close:2013, Morzinski2014, Morzinski2016} system's visible light camera \citep[VisAO,][]{Males2013diss, Males2014a} in H$\alpha$ Simultaneous Differential Imaging (SDI) mode. The data were acquired between 2013 and 2018 as part of the Giant Accreting Protoplanet Survey \citep[GAPlanetS][]{folletteinprep}, a search for protoplanets inside of the gaps of transitional disk host stars. GAPlanetS data are processed with a custom IDL pipeline, as described in detail in \citep{follette2017}. In brief, they are dark-subtracted and divided by a flat field image, which is generally acquired once per observing semester. This corrects primarily for attenuation of light by near-focus dust spots on the CCD window, as the VisAO CCD is otherwise flat to within 1\%. The images are centered and aligned using Fourier cross-correlation, separated into H$\alpha$ ($\lambda_c=656nm, \Delta\lambda=6nm$) and Continuum ($\lambda_c=642nm, \Delta\lambda=6nm$) channels (acquired simultaneously on the detector), and cropped to a 451 pixel ($\sim3.5"$) square. Images with cosmic rays within 50 pixels of the central star are removed from the image cube before analysis.

This work focuses on the transitional disk system HD 142527, which has a known { $0.26^{+0.16}_{-0.14}~\mathrm{M_\odot}$ \citep{Claudi2019} companion \citep[HD 142527 B]{Close2008, biller2012} at separations ranging from $12.38 - 22.89$ AU} \citep{Balmer2022}. In particular, we are interested in HD 142527 epochs where the companion is detectable with a range of KLIP parameters. These robust detections provide a stable ``training'' set of data for optimization. This was not the case for the GAPlanetS data collected on 10 February 2017, for which the companion was undetectable at SNR greater than 3 using any KLIP parameters. This is likely due to the limited rotation of this dataset (16.1$^{\circ}$), and the companion's tight separation ($44.29\pm2.57$ mas \citep{Balmer2022}).

 {Seeing data indicated sub-arcsecond conditions for all datasets for which it was available, however the site seeing was non-operational during two of the epochs. The FWHM of the datasets ranges from 4-5.5 pixels (32-44mas) with an average of 4.8 pixels or 38mas, and on-sky rotation varies from 34.8-117.4$^o$ with an average of 74$^o$.} The properties of the datasets used are described in Table \ref{Tab:obs}. Astrometry and photometry of the HD142527B companion from these datasets are discussed in detail in \citet{Balmer2022}.

\subsection{Dataset Selection}

Multiwavelength imaging of targets with known companions is ideal for testing optimization of the \texttt{pyKLIP} algorithm. Wavelengths where the companion is fainter or not visible can be leveraged as ``clean" images into which synthetic companions/planets can be inserted and optimized. Then, the efficacy of a parameter optimization approach can be evaluated by applying it to target wavelengths. 


In exploring data-driven approaches to KLIP optimization for GAPlanetS data, we focus on optimization of simulated planets injected at a range of separations into the Continuum images for HD142527~B datasets. We then apply this approach to the HD142527 H$\alpha$ datasets to test whether the approach results in robust single epoch H$\alpha$ recovery of the known accreting companion in these datasets. In Section \ref{results}, we discuss the SNR penalty of optimizing on injected planets injected into Continuum images rather than directly on H$\alpha$ images. 

We note that we have focused our approach in this work on detection of planets at H$\alpha$ alone, and have not applied Simultaneous Differential Imaging (SDI) as a part of our optimization approach {i.e. We used continuum imagery to inject and optimize false planets, but did not combine H$\alpha$ and continuum images before or after PSF subtraction, treating them as wholly separate datasets}. SDI results are reported in the GAPlanetS Survey paper \citet{folletteinprep} released in conjunction with this work. 

Although our analysis of this proposed approach to optimization relies on one object, the companion exists at a wide range of separations ($\sim$40-80mas) across the 2013-2018 time baseline and the datasets are of widely varying quality.

\subsection{Pre-KLIP ``Data Quality Cuts"} \label{dataqualitycuts}

In order to reduce the computational scope of the optimization problem, we implemented ``data quality cuts" prior to KLIP optimization as described in detail in \citet{folletteinprep}. In short, we fit the stellar PSF (or ghost in the case of saturated images) with a Moffat profile and extracted the peak value of this fit for every image in an image sequence. Using the peak value as a proxy for the ``quality" of a given image, we then culled the data by discarding the images with the lowest peaks. We compared contrast curves (computed under a conservative set of KLIP parameters and with a highpass filter of width of 0.5$\times$FWHM) for the full image sequence to those with 5, 10, 20, 30, 40, 50, 60, 70, 80, and 90\% of the lowest quality images discarded. We then chose a data quality cut by eye, balancing the overall contrast in the inner and outer regions of the image, generally adopting the cut with the highest achieved contrast (lowest contrast curve) in the inner 0$\farcs$1-0$\farcs$25 unless that cut was substantially worse in the outer regions of the image. The selected data quality cuts are listed in Table \ref{Tab:fakes}. 

This technique, which builds on the principle of ``Lucky Imaging" \citep{Fried1978}, is a remarkably nuanced one. The ``answer" for the optimal data quality cut for a given image set appears to vary with location within the image. We do not explore it in detail in this work, but we do note here that ultimately, this parameter is likely also an important consideration for future optimization work and should in principle be optimized together with \texttt{KLIP} parameters.

\subsection{Synthetic Planet Injection} \label{falseinject}

The cornerstone of our proposed optimization approach is the assertion that KLIP optimization should be done on \textit{simulated} companions to avoid cognitive biases in parameter selection. Therefore, another critical pre-KLIP step in our approach is injecting synthetic planets into the Continuum images.

Among other considerations in planet injection is the region of the post-processed image in which to optimize detections. In this work, we have opted to focus on the region inside of the Adaptive Optics (AO) system's control radius for planet insertion and recovery. This region (also known as the `dark hole") is in most cases equivalent to or larger than the size of the cleared central cavities of GAPlanetS transitional disks, which is the region most likely to host detectable accreting protoplanets. In the case of HD142527, however, the cavity is very large \citep[$\sim1 \arcsec$][]{Avenhaus2014}. It is likely that optimization of companions in the outer portion of the HD142527 cavity would yield different choices for optimal KLIP parameters than those reported here.

Injecting planets into raw data and recovering them at a range of separations and position angles is a common method for quantifying azimuthal variation in recovered astrometry and photometry due to PSF asymmetries (e.g.\cite{Wagner2018}). It also serves as a means to quantify the degree of flux attenuation (``throughput") introduced by the KLIP algorithm as a function of separation, which is needed to determine detection limits \citep[][]{Mawet2019}. 

This technique of inserting synthetic planet signals is also a potential tool for optimization. We hypothesize that if synthetic planets injected into Continuum images are successful at mimicking real planets, their optimal KLIP parameters should result in high-quality recoveries of real companions in H$\alpha$ datasets. This hypothesis is examined in detail in Section \ref{truevfake}.  

To this end, we injected planetary signals into the raw Continuum image sequences (culled according to the chosen ``data quality" cuts). The Continuum images were chosen in order to reduce the influence of the companion on the results, though we note that it is recoverable at moderate SNR \citep[SNR=2-6][]{Balmer2022} in Continuum wavelengths. In the case of HD 142527, we injected planets away from the companion to mitigate its effects; which are then limited to an inflation of the noise statistics at the companion's separation. In this way, the HD142527 data mimics a case where optimization is being done on injected planets in the vicinity of real, previously unknown, planets. 

Synthetic planets are created by scaling the images of the central star (or ghost in the case of saturated data) to a desired contrast and injecting them into the raw images. The PSF of the injected planets is constructed from the stellar PSF image-by-image, so intrinsic variation in the stellar PSF is also captured in the injected planetary PSF (as would be the case with real planets). Contrasts of these injected planets are computed under a single set of KLIP parameters (\texttt{annuli}=5, \texttt{movement}=2) that {ad-hoc experimentation with MagAO data indicates will result in robust recovery of most companions.} These injected planet contrasts were iterated upon until the average SNR across 5, 10, 20, and 50 KL modes was in the range of 6.5-7.5. We injected as many planets as would fit between the inner working angle (PSF FWHM) and the control radius separated by 1 FWHM radially. More specifically: 

\begin{equation}
    \mathrm{N_{injected}} = \frac{\mathrm{Control\ Radius} - \mathrm{IWA}}{\mathrm{FWHM}} 
\end{equation}

{The position angles (PAs) of these fake injected planets were assigned such that the first planet would be placed at a PA of 0, and each subsequent planet would be advanced by a PA of 85 degrees, probing different azimuthal regions of the PSF.}

Figure \ref{fig:schematic}, Step 2 visualizes the stage in our reduction process where synthetic planet injection is done, and shows the resulting fake planets. Table \ref{Tab:fakes} shows the contrasts and number of fake injected planets in each dataset. 

\begin{deluxetable*}{cccc}
    \tablehead{
    \colhead{{Date}}  & \colhead{{Cut}} &  \colhead{{Fake Contrast}} & \colhead{$\mathrm{N_{injected}}$}
    }
    \startdata
    11 April 13  & 10 & 0.01 & 8 \\
    8 April 14  & 0 & 0.01 & 5 \\
    15 May 15  & 50 & 0.01 & 5 \\
    16 May 15  & 80 & 0.05 & 5 \\
    18 May 15  & 0 & 0.01 & 4 \\
    27 April 18  & 0 & 0.05 & 8 \\
    \enddata
    \label{Tab:fakes}
    \caption{Injected Planet Parameters. Fake Contrast is the contrast of injected fake planets used to compute contrast curves(this same contrast is used to inject synthetic planets for pyKLIP-PE optimization). $\mathrm{N_{injected}}$ is the number of injected planets between the IWA and control radius used to compute the optimal parameters. }
\end{deluxetable*}

\section{Optimization Methods} \label{image_processing}

The goal of this work is to develop a method for selecting KLIP parameter combinations that is 1) robust to false positives and 2) leads to high-quality recoveries of known objects. At the same time, we aim to develop a method that does not rely on the planetary signal itself for optimization. This approach is critical both in developing strategies for untargeted/uninformed exoplanet searches, where the location(s) of planet(s) are not known \textit{a priori}, and in reducing cognitive biases in parameter optimization. 

In other words, we should be able to apply these techniques to robustly recover planet signal without prior information about the planet's existence. In doing so, a first key question is what the metrics for a `good' planet recovery should entail. A second is how we should combine these metrics to refine an optimization approach. We apply several techniques to this problem, as outlined below. 




\subsection{Image Quality Metrics} \label{imagequalmetrics}

In order to make decisions among possible \texttt{pyKLIP} parameter values, one or more metrics for the``quality" of the signal extraction are required. Due to the nuanced nature of post-processed image quality, we chose to utilize three measurables in our approach to optimization: Signal-to-Noise Ratio (SNR) maps, Contrast Curves, and False Positive Thresholds.

\subsubsection{Peak/Average SNR}

A common metric for a high quality recovery is a planet's signal-to-noise ratio (SNR) i.e., its signal should be at least $3$ to $5 \sigma$ above the noise level, where the noise level is the standard deviation of background pixels at that separation. 

In order to compute SNR Maps, we mask a wedge-shaped region with a radial width of the PSF FWHM and an azimuthal width of 15$^{\circ}$ on either side of each planet, which masks the planet itself and all or most of the ``self-subtraction lobes" that extend azimuthally from its location. We then estimate the noise at each separation by computing the standard deviation of the non-masked (noise) pixels in 1 pixel \texttt{annuli} and apply a statistical correction following \citet{mawet2014} to account for the small number of independent noise samples near the central star. 

The classic measure of the quality of a High Contrast Imaging detection is the maximum single pixel value in an SNR Map (which we will call ``peak SNR" hereafter) of a recovered companion. However, we also extract the {average} SNR of the positive pixels under our planet masks as an alternative, potentially more robust, measure of the detection quality. It should be noted that, in our case, SNR is measured on post-processed images that have been subject to a highpass filter, and that the final post-processed images were smoothed with a 1 pixel Gaussian kernel. 


\subsubsection{Contrast}

Another common metric for the quality of a high-contrast imaging reduction is the achieved planet/star contrast limit. The optimal reduction by this metric should have the best planet/star contrast (lowest contrast value), allowing for the recovery of the faintest objects. 

As is standard in the field, we compute contrast by 1) measuring the 5$\sigma$ noise at a given location and 2) calibrating that noise to correct for the algorithm throughput. Throughput is computed as the ratio of a planet's true (injected) brightness to its post-processed (recovered) brightness. 
The ``raw'' 5$\sigma$ noise level divided by the throughput yields a limiting brightness for planets to be recovered at 5$\sigma$ at a given location in the image. 

\subsubsection{False Positive Fraction}
High quality HCI reductions are also those in which all signals meeting the canonical SNR threshold are true signals and not false positives. On the assumption that there are no additional true signals in our images, we consider false positive pixels to be those with values in the post-processed maps with SNRs above 5$\sigma$ that are not at the location of the known or injected companion(s). We count the number of pixels between the IWA and control radius that meet this threshold, excluding those under the planetary mask(s). We note that more nuanced approaches to false positive estimation are possible. Isolated single pixels with high values, for example, are less likely to be mistaken for a companion than clusters of pixels with high values. Previously unknown \textit{real} companions that have not been masked and are visible at the optimization wavelength will also influence this value, as is the case in all of our datasets. Future work should implement a more nuanced version of this metric, perhaps also incorporating a forward modeling approach (e.g. \cite{Ruffio_2017}).

\subsubsection{Neighbor Quality}

`Neighbour Quality' metrics are created by smoothing the peak and average SNR metrics in movement/annuli space by a Gaussian with a FWHM of 3 pixels. They serve to create a measure of the quality of \textit{neighboring} parameters in movement/annuli space. In other words, a high SNR value that is an outlier in a region of the annuli/movement heatmap is penalized by the low SNRs of neighboring values. Parameter regions that are stably high, such that small changes in movement/annuli values result in minimal changes in SNR, are given additional weight. A neighbour quality metric is created for both Peak SNR and Average SNR


    
     



\subsection{pyKLIP-PE} \label{pyKLIPPE}

The final tool in our approach to optimization is the \texttt{pyKLIP} Parameter Explorer (\texttt{pyKLIP-PE}). This coarse grid search algorithm runs \texttt{pyKLIP} for every combination of specified \texttt{movement} and \texttt{annuli} parameter values for a specified range of \texttt{KL mode} values, and returns all of the image quality metrics specified in Section \ref{imagequalmetrics}. 

The output for each dataset is a 5D cube with the following dimensions: \texttt{annuli}, \texttt{movement}, \texttt{numbasis} (KL Mode), planet, and image quality metric. In this work, we explore \texttt{annuli} values of 1 to 25, \texttt{movement} values of 0 to 24, and a KL basis set of \texttt{numbasis=[1,2,3,4,5,10,20,50,100]}. The maxima of the annuli and movement ranges are meant to limit the computational time of each run of the algorithm, but can be expanded or reduced as necessary. KL mode values 1-5 were selected due to the fact that they capture the most frequently occurring PSF modes. KL Modes 10, 20, 50 and 100 were subsequently chosen in an attempt to capture a range of complexities gained by adding more KL Modes.

To produce uniformity among datasets, we standardized each of the image quality metrics to take values between $\sim$0 and 1 by subtracting the minimum value and dividing my the maximum - minimum value so that 1 is the ``best" combination of parameters and 0 is the ``worst". In each case, this normalization is computed in \texttt{movement}/\texttt{annuli} space across KL modes so that for each planet, metric, and KL mode combination the "optimal" answer may vary across these parameters. 

For contrast, we compute an image quality metric in log contrast space that ranges from 0 to 1 as follows:
$$\frac{ log_{10}(C)  -  log_{10}(C_{min})}{ log_{10}(C_{max}) - log_{10}(C_{min}) } $$
where, $C$ is the measured 5$\sigma$ contrast for each injected planet, KL Mode, annuli, and movement combination, and $C_{min}$ and $C_{max}$ are the minimum and maximum 5$\sigma$ contrast values across all KL mode, annuli, movement combinations for that planet.

We then convert the false positive pixel count to a normalized metric For each pixel exceeding the 5$\sigma$ threshold inside the control radius for each KL Mode, annuli, and movement combination, we 1) Subtract its minimum value (generally 0),  2)  Divide by the difference between its maximum and minimum value across all KL modes. 3) Subtract this value from 1. This creates a metric for which the \texttt{movement} and \texttt{annuli} parameters with the highest number of false positives are assigned a value of 0, and those with the lowest are assigned values near the maximum of 1.

\begin{figure*}
    \begin{center}
        \includegraphics[width=\textwidth, height=\linewidth, keepaspectratio]{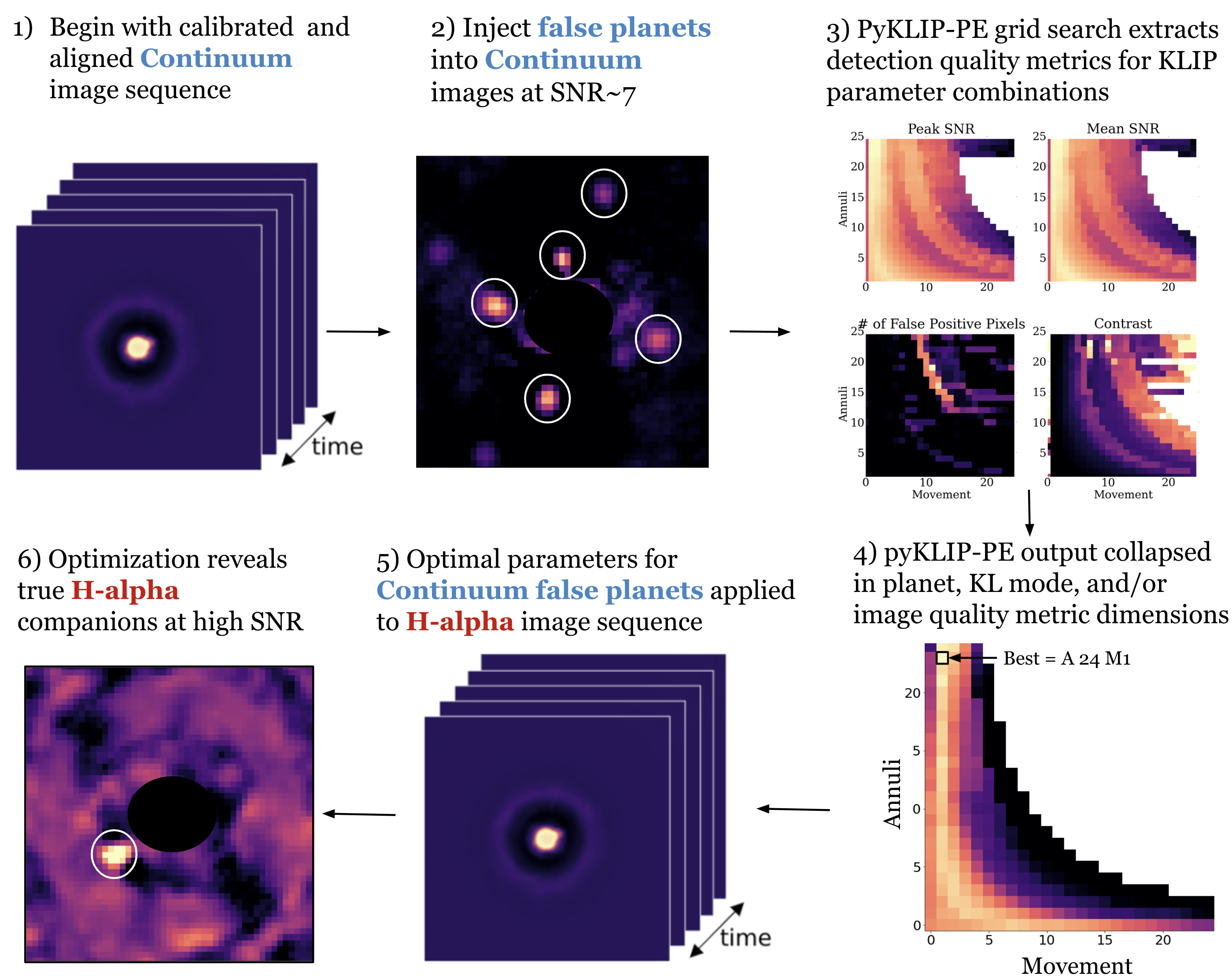}
    \caption{A schematic representation of the optimization process applied in this work, with images drawn from optimization of the HD142527 11Apr13 dataset.}
    \label{fig:schematic}
    \end{center}
\end{figure*}

A schematic depicting the various stages of the optimization process outlined in this section is shown in Figure \ref{fig:schematic}.



Note that the nature of the 5D \texttt{pyKLIP-PE} output parameter space is such that it can be collapsed in many ways to select the ``best" choice of \texttt{annuli}, \texttt{movement}, and \texttt{numbasis} parameters for a given dataset. In Section \ref{results} we explore several ways in which we chose to collapse these cubes to select a ``best" parameter choice, though we note that there are many other possibilities that remain to be explored.

Once optimal parameters have been selected for synthetic planets injected into the Continuum images, we apply those parameters to the unaltered (no simulated planets) $H \alpha$ images to test the ability of this optimization method to robustly recover HD 142527 B at H$\alpha$ (see Table \ref{Tab:obs} for details of these observations). At this stage, we apply our optimized parameters to the H$\alpha$ images only, and do not engage in SDI reductions. 

\subsection{\texttt{pyKLIP-PE} Structure}
Because the movement parameter is applied at the center of each annular region and the width of these regions vary with the value of the annuli parameter, certain combinations of annuli and movement lead to equivalent sets of reference images (equivalent rotational masks) and broadly similar reductions. For this reason, parameter exploration outputs frequently show diagonal structure from regions with a large number of annuli and low movement values to fewer annuli with larger movement values. 

A further feature of the parameter explorer heatmaps are the jagged structures along the right-hand (high movement value) side. These occur when the synthetic planets are "passed" by the annular zone boundaries and move from being at the inner edge of an annulus whose center is at larger radii to the outer edge of an annulus whose center is at smaller radii. This shift to a new annulus allows for greater movement values to be applied before the algorithm runs out of reference images in the new zone. 

Unphysical contrast values are also assigned \texttt{NaN} values, leading to a further source of white pixels in each parameter explorer.

\subsection{Fixed Parameters}

In order to make the computational time for GAPlanetS datasets tractable, we chose to limit the \texttt{pyKLIP-PE} grid search to only the \texttt{annuli}, \texttt{movement}, and \texttt{numbasis} KLIP parameters. We did, however, explore in a less systematic way the effect of the highpass filter (\texttt{highpass}) and inner working angle (\texttt{IWA}) values on post-processed images, and arrived at what we deemed to be reasonable fixed choices for the values of these additional influential KLIP input parameters. 

\subsubsection{Highpass} 

\begin{figure*}
    \begin{center}
        \includegraphics[width=\textwidth, height=\linewidth, keepaspectratio]{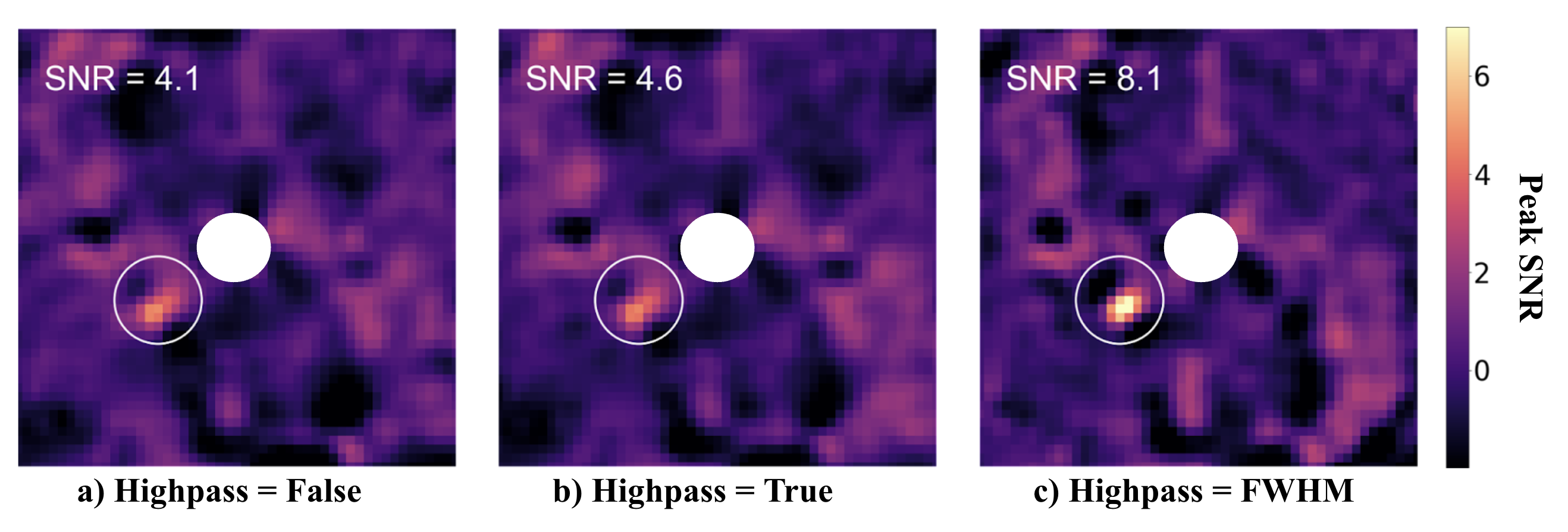}
    \caption{SNR Maps showing the effect of highpass filtering on the HD142527 11 April, 2013 dataset. The leftmost SNR Map shows no highpass filtering (\texttt{highpass} = False), and has a peak companion SNR of 4.1. The middle SNR Map shows a default highpass filter width of $0.1 \times$ the image size (\texttt{highpass} = True) and has a slightly higher Peak SNR of 4.6. The rightmost SNR Map shows a more aggressive highpass filter width of  $1 \times$ FWHM and has a Peak SNR of 8.1.}
    \label{fig:highpass_new}
    \end{center}
\end{figure*}

In the case of the Fourier highpass filter width parameter \texttt{highpass}, we explored setting it to values of ``False", ``True" (filter size = image size / 10), and the data FWHM for several HD142527 datasets. We find that the application of a an aggressive highpass filter with a size near the stellar FWHM has a positive effect on the image quality when compared to images with no highpass filter (\texttt{highpass}=False) or a conservative highpass filter width (\texttt{highpass}=True). Example SNR maps showing various highpass filter widths on the HD142527 B 11 April, 2013 dataset are displayed in Figure \ref{fig:highpass_new}. With no highpass filtering, the detection has an SNR of 4.1. With highpass filter set to ``True", a width of 1/10th the image size, the SNR improves minimally to a value of 4.5. However, when we set the width to 1 FWHM, the SNR improved by nearly a factor of 2 to 8.1$\sigma$.  

The effect of each filter size on image and detection quality were further explored with Receiver Operating Characteristic (ROC) Curves, and Contrast Curves for the HD 142527 datasets. We explored  0.5, 1, or 1.5 times the FWHM, but found no significant difference between these filter widths across datasets. Further analyses need to be done to understand the nuanced trade-offs of highpass values in this regime.

\subsubsection{Inner Working Angle} 

In the case of the Inner Working Angle parameter \texttt{IWA}, MagAO data are non-coronagraphic, so this parameter can in principle be set to zero.  The cleared central cavities of GAPlanetS targets are generally $<$0$\farcs$1 in radius, beneath the IWA of many coronagraphic HCI instuments, and the known companions in their gaps are very tightly separated: smaller IWAs leave more area for planet detection in this region. However, for initial \texttt{pyKLIP} parameters \texttt{annuli = 5}, \texttt{movement = 2}, \texttt{KL = 10}, we found that in 5 out of the 6 datasets (5/6), setting the IWA = FWHM instead of IWA = 0 moderately improved recovered SNR values, with an average improvement of 0.62 $\sigma$.

We therefore elected to balance these considerations by setting \texttt{IWA} to a fixed value of either the median FWHM of the dataset or, in the case of saturated data, the saturation radius. We chose these values because we would not be able to separate the light of companions from the starlight regardless of their intrinsic brightness at these separations. The range of IWA values for the datasets considered here spans 3-8 pixels.

We note that, because the KLIP \texttt{annuli} parameter divides the space between the IWA and Outer Working Angle (OWA, in this case the image boundary) into evenly spaced annular zones, modifying the IWA from dataset to dataset results in a small variation in the ``meaning" of the \texttt{annuli} parameter values, in the sense that the same number of annular zones may have slightly different annular widths in pixels if the OWA-IWA range is varied. However, this difference in annuli widths ranges from just 5 pixels for the lowest \texttt{annuli} value (1) to under 1 pixel for the highest (25). Similarly, because the \texttt{movement} parameter is defined relative to the center of each annulus, this results in a very small variation ($<1^{\circ}$) in the size of the rotational mask corresponding to each \texttt{movement} value. 



Although we chose to fix the \texttt{IWA}, and \texttt{highpass} \texttt{pyKLIP} parameters in this investigation, with greater computational resources and/or for smaller single datasets, we strongly recommend optimizing them as well.


56
\section{Optimization Analysis and Results}\label{results}

In an uninformed search for planets, we do not know \textit{a priori} whether companions are present or at what separations and position angles they might appear in our datasets. For this reason, we've chosen to adopt an approach that averages over a number of planets injected at different separations and position angles in our Continuum images, effectively optimizing over a region of interest rather than at a single azimuthal and radial location. 
 
 
The HD 142527 datasets, with a known companion at a range of separations across the time baseline of our observations, serve as an ideal test case for this general methodology. If the Continuum, multi-planet optimized, set of KLIP parameters is robust, then it should successfully recover true H$\alpha$ companion signals at any location within the optimization region at reasonably high SNR. This SNR may not be the highest achievable, since the companion itself was not the target of the optimization. However, the method does not rely on the reality of an apparent companion signal, and therefore is more robust to cognitive biases, particularly in the case of low SNR detections. It also mitigates the risk of artificially inflating the SNR of a companion signal by optimizing on a single localized combination of companion signal and speckle field. In this section, we report the results of KLIP ADI optimization based on five detection metrics (see Section \ref{imagequalmetrics} for a description of these metrics) in Section \ref{optworks}. We then compare the results of these optimizations to real planet optimization in Section \ref{truevfake}.

\subsection{Generic vs Optimized Parameter SNR Maps} \label{optworks}

\begin{figure*}
    \begin{center}
        \includegraphics[width=\textwidth, height=\linewidth, keepaspectratio]{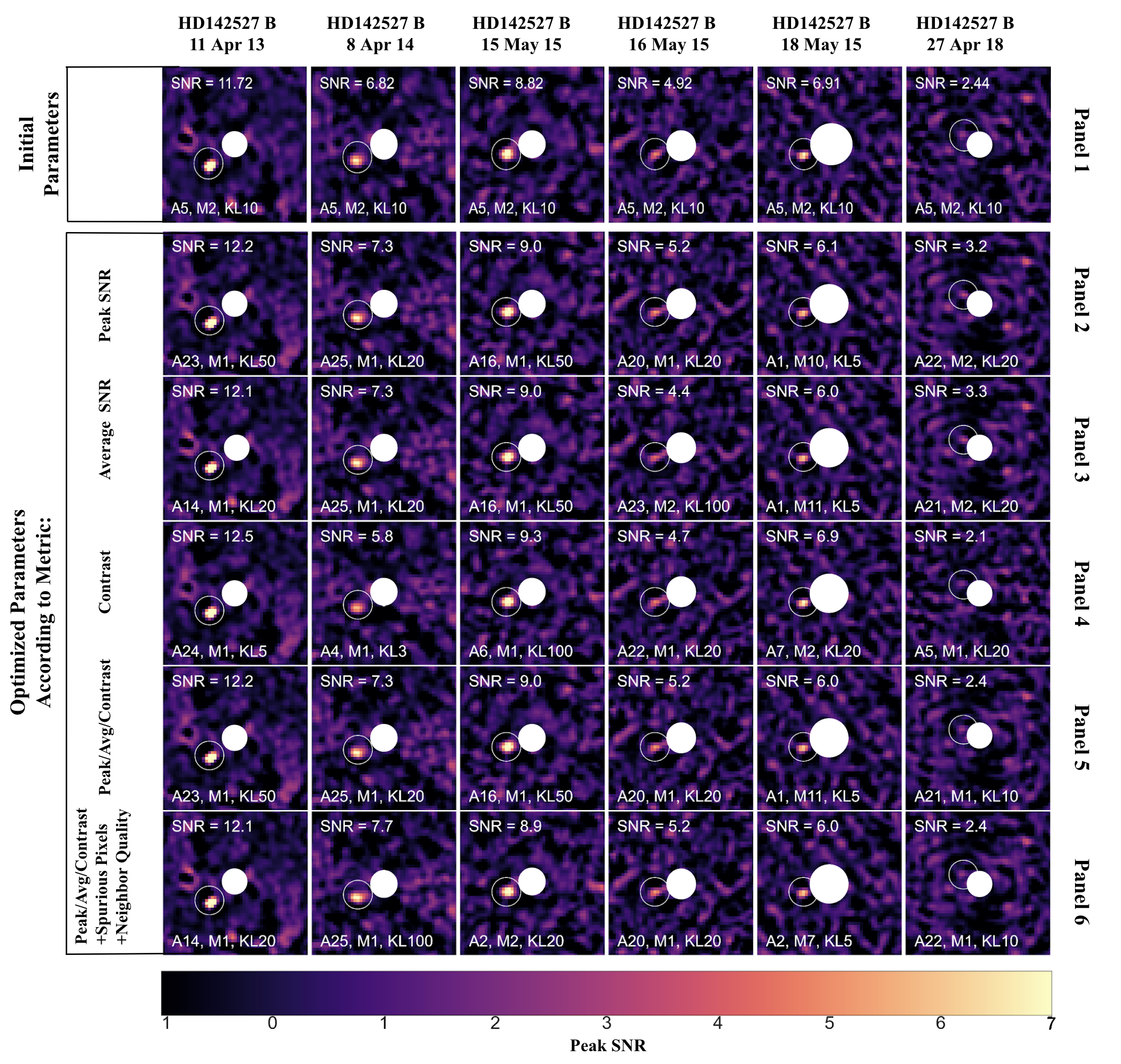}
    \caption{Optimization results for six HD 142527 datasets in which the companion is recovered. The top panel shows SNR Maps of KLIP reductions using a single set of fixed KLIP parameters for all datasets: \texttt{annuli} = 5, \texttt{movement} = 2, \texttt{KL} = 10. Panels two, three and four show SNR maps of H$\alpha$ KLIP reductions with optimal parameters identified for injected Continuum planets under the \texttt{pyKLIP-PE} image quality metrics Peak SNR (Panel 2), Average SNR (Panel 3), Contrast (Panel 4) \textit{individually}. Panel 5 shows the optimal parameters selected by the \textit{combination} of these three metrics. Panel six shows the Peak SNR, average SNR, contrast combination along with the false positive pixel and neighbor quality metrics.  Note that the best parameters based on false positive pixels are not shown individually because they are nearly degenerate for most datasets. In each of the six epochs, \texttt{pyKLIP-PE} was able to recover the real planet at a higher SNR than the generic reductions under each individual metric in 22/30 cases. A Gaussian smoothing was applied to these data with a standard deviation of 1 before computing the SNR maps.}
    \label{fig:allmetrics}
    \end{center}
\end{figure*}

In many direct imaging surveys, KLIP reductions are done with the same set of parameters across all datasets. We mimicked this ``survey'' strategy by selecting a single, generic set of KLIP parameters and applying them to all HD 142527 datasets for comparison against our optimized reductions. Previous experience with these GAPlanetS datasets led us to choose \texttt{annuli} = 5, \texttt{movement} = 2, \texttt{KL Mode} = 10, which reveals most sources in our data at an SNR greater than 2. KLIP reductions with these initial parameters are shown in the top panel of Figure \ref{fig:allmetrics}. 

We then ran \texttt{pyKLIP-PE} on Continuum planets injected into our six HD 142527 Continuum datasets, with the synthetic planets injected as described in Section \ref{falseinject}. From the \texttt{pyKLIP-PE} output, we first extracted optimum \texttt{movement}, \texttt{annuli} and \texttt{KL mode} parameters for each of the three main image quality metrics individually, namely: Peak SNR, Average SNR, and Contrast (see Section \ref{image_processing} for a description of how these metrics are computed). These are shown in the 2nd, 3rd and 4th panels of Figure \ref{fig:allmetrics}. We then extracted the optimal parameters from the combination of these three metrics, as shown in the 5th panel of Figure \ref{fig:allmetrics}. Finally, we incorporated the False Positive Fraction and Neighbor Quality metrics described in Section \ref{image_processing}, and combined all six image quality metrics, shown in the bottom panel of Figure \ref{fig:allmetrics}. The Peak SNR of each reduction is reported as is standard in direct image.

Figure \ref{fig:allmetrics} is therefore a gallery of KLIP post-processed H$\alpha$ images whose parameters were selected to maximize the recovery of Continuum injected planets under each metric. Note that the false positive pixel metric is not shown here individually because it is nearly degenerate and does not have a single optimal value. It is therefore most beneficial when combined with other metrics as a means of excluding bad parameter choices. 

Of the 30 optimized companion detections depicted in Figure \ref{fig:allmetrics} (6 datasets x 5 metric combinations), 22 produced a Peak SNR improvement of up to $1.2 \sigma$ relative to reductions with generic KLIP parameters (\texttt{annuli} = 5, \texttt{movement} = 2, \texttt{KL Mode} = 10). For five out of the six datasets, there is at least one metric choice that improved detection SNR over the generic with an average improvement of 0.6. In the remaining dataset, the maximum detection SNR selected by the metrics is equivalent to that of the generic parameters. However, even in cases where the optimal false planet parameters for a particular image quality metric lowered the SNR relative to the generic parameter choices, there was still a clear detection of the companion using all metrics. In the case of the most difficult dataset (27 April 2018), where the companion is separated by only 5 pixels from the central star, the generic reduction could be classified as a non-detection, while the \texttt{pyKLIP-PE} reductions maximizing Peak SNR and Average SNR reveals the object at an SNR of $\sim 3.2 \sigma$. Therefore, we find that in an uninformed or untargeted planet search of the HD 142527 system, \texttt{pyKLIP-PE} would recover the companion at an SNR greater than 3 in all epochs using a number of metric combinations.

The Peak SNR metric optimization produced the best detections in five out of the six datasets analyzed. However, its optimizations were not significantly better than any other single metric. We also find that for a given dataset, multiple metrics tend to converge on the same parameters (e.g. in 8 April, 2014, four out of our five metrics were maximized at \texttt{annuli = 25}, \texttt{movement = 1}. In Section \ref{truevfake}, we explore the effects of combining multiple metrics, as well as the effect of averaging across KL modes. 

\subsection{Comparison of False and True Companion Optimization} \label{truevfake}

In the previous section, we show a gallery of KLIP H$\alpha$ reductions with parameters chosen to maximize various combinations of the image quality metrics: Peak SNR, Average SNR, Contrast, False Positive Pixels, and Neighbor Quality. In each case, we ran the \texttt{pyKLIP-PE} optimizer on fake planets injected into the Continuum data, and applied the selected optimal parameters to H$\alpha$ images to reveal the real companion. However, in a case such as this -- where the existence of the companion is well established -- a direct H$\alpha$ optimization result could in principle be used in lieu of the Continuum injected planet output to maximize the recovered signal of the planet. We expect that parameters selected in this manner will better reveal the companion than parameters chosen via fake planet injection. Therefore, an important test of the \texttt{pyKLIP-PE}'s efficacy is its level of agreement between fake and real planet optimization. 

In order to estimate the level of agreement between injected and true planet optimization, we first reduced the dimensionality of the \texttt{pyKLIP-PE} output. This output consists of twenty-five movement values, twenty-five annuli values, nine KL modes (1,2,3,4,5,10,20,50,100), six parameter quality metrics (Peak SNR, Peak SNR Neighbor Quality, Average SNR, Average SNR Neighbor Quality, Spurious Pixel Count, and Contrast), and 4-8 injected planets for each dataset. We begin by averaging across the Continuum planets injected at various PAs and separations between the IWA and control radius (the ``optimization region"), which allows for direct comparison with the single companion in H$\alpha$. We then collapsed the \texttt{pyKLIP-PE} output for each dataset according to every possible combination of image quality metrics and KL modes. The result is an aggregate data quality (ADQ) map for each possible metric collapse scenario. Since these aggregate maps are combinations of the normalized image quality metrics, they have a range of possible maximum values equal to the number of image quality metrics that have been combined. We re-normalize them by subtracting the 10th percentile value, and dividing by the 90th percentile, creating a parameter quality map where most values lie between 0 and 1. We do not explore weighting of the image quality metrics in this work, and combine them as a simple linear combination of the individual metrics, but the application of weighting coefficients is likely a fruitful avenue for future exploration.

\begin{figure*}
    \centering
    \includegraphics[width=5.8in]{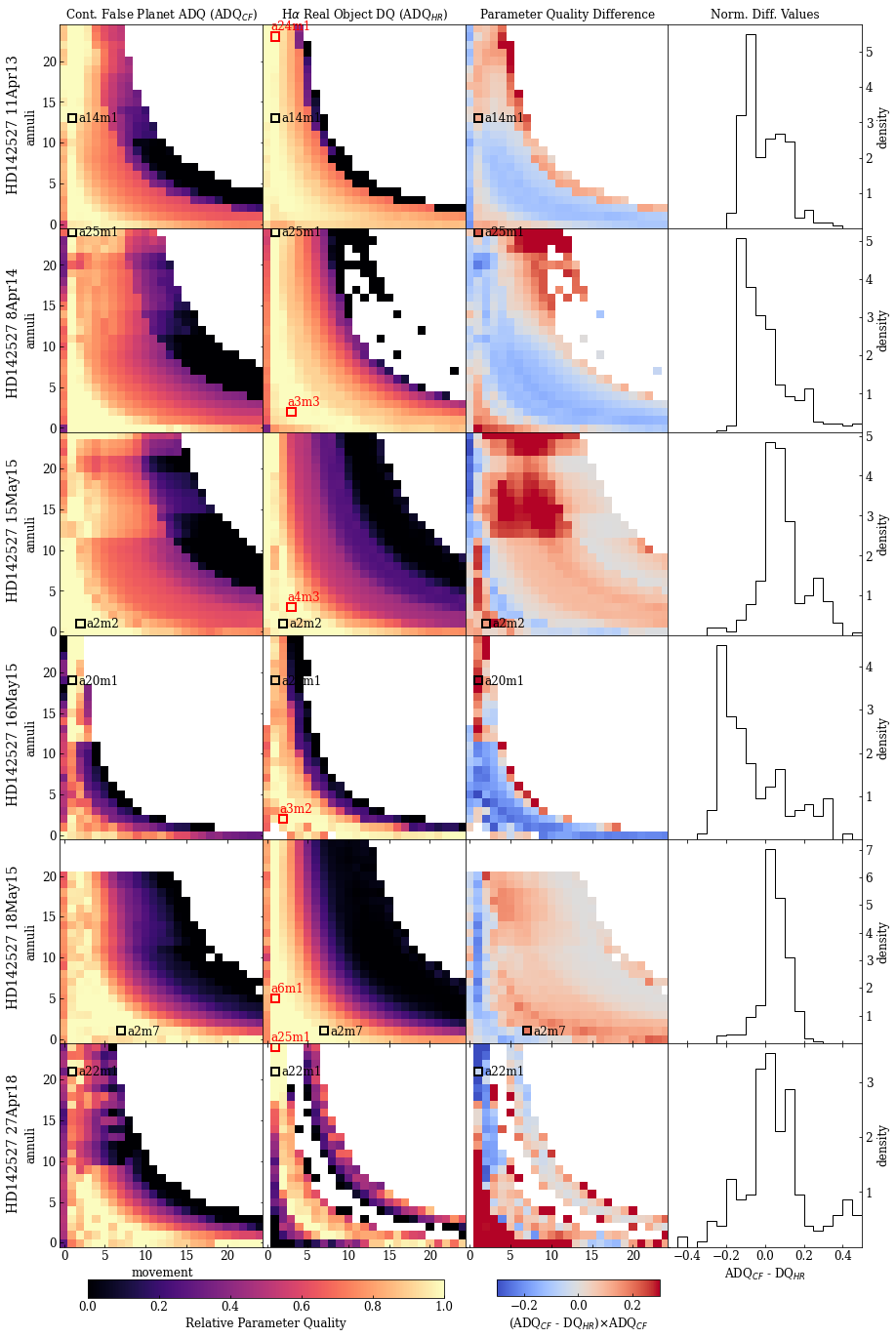}
    \caption{Maps of \texttt{pyKLIP-PE} output for Continuum injected (CF) planets (first column) and the H$\alpha$ real (HR) companion (second column), for each HD142527B dataset, normalized so that the 10th-90th percentiles in the {aggregate data quality (ADQ) metric span the range 0 to 1}. The difference between the two normalized maps is shown in column 3, where red values indicate parameter combinations that are more strongly favored for false planets, and blue indicates the opposite. Shown in the rightmost column are histograms of the difference maps. The highest relative parameter quality annuli / movement combinations are highlighted in red in the two leftmost columns.}
    \label{fig:TFplanets}
\end{figure*}

{In this work we use these H$\alpha$ optimization maps only to test the structural consistency between Continuum injected planet aggregate data quality (ADQ) maps under each metric collapse scenario with those of the true H$\alpha$ companion.} To quantify the consistency between the two parameter quality maps (Continuum injected planets and true H$\alpha$ companion), we subtract the map for the true H$\alpha$ companion from that of the Continuum injected planets. This creates a visualization of the difference in the \textit{structure} of the parameter qualities (see Figure \ref{fig:TFplanets}). We use summary statistics of these difference maps to analyze the relative merits of different collapse methods and metric combinations, as well as qualitative metrics such as the appearance of the final difference maps in the low movement regions where we expect the results to be most stable. This includes searching for the combinations that minimize the differences in the structure of \texttt{pyKLIP-PE} parameter space, which are the values where the true and false companion results are most closely aligned. 

One pattern present in the H$\alpha$ aggregate data quality (ADQ) maps (shown in the left hand panels of Figure \ref{fig:TFplanets}) is that they often exhibit two distinct peaks. There is commonly a region of stably optimal parameter space at low \texttt{annuli} and another at high \texttt{annuli} values. Movement space shows fewer clear patterns in the false vs. true companion residual maps. However, movement values of zero, equivalent to no rotational mask, are generally unstable in most image metrics, so we have chosen to exclude them in this analysis. 

We initially sought to identify the metric or sum of metrics and KL modes that minimized the difference between the real and injected planet optimizations. In principle, a collapse method that mimics the structure of the true companion parameter explorer will have a low standard deviation in the difference map, while maintaining a sum, median, and difference at the Continuum peak near zero. Therefore, we utilized distributions of the following values from the difference maps across all possible metric and KL combinations for all six HD142527 datasets: the median, standard deviation, and sum of the difference maps (1-3), and the value of the difference map at the location of the Continuum injected planet peak (4), which quantifies the relative penalty of optimizing on the combination of Continuum planets in lieu of the single known companion. 

We isolated all metric combinations for which the standard deviation of the false-true difference map was among the lowest X\%, and where the sum, median, and difference at the peak of that map were among the X\% closest to the mean, where X is a value that we varied to get a sense for the patterns in these parameters under the assumption that, with the small number of datasets under consideration, patterns in ``good" metrics would be more informative/universal than the single ``best" metric. 

\begin{figure*}
    \begin{center}
        \includegraphics[width=\textwidth, height=\linewidth, keepaspectratio]{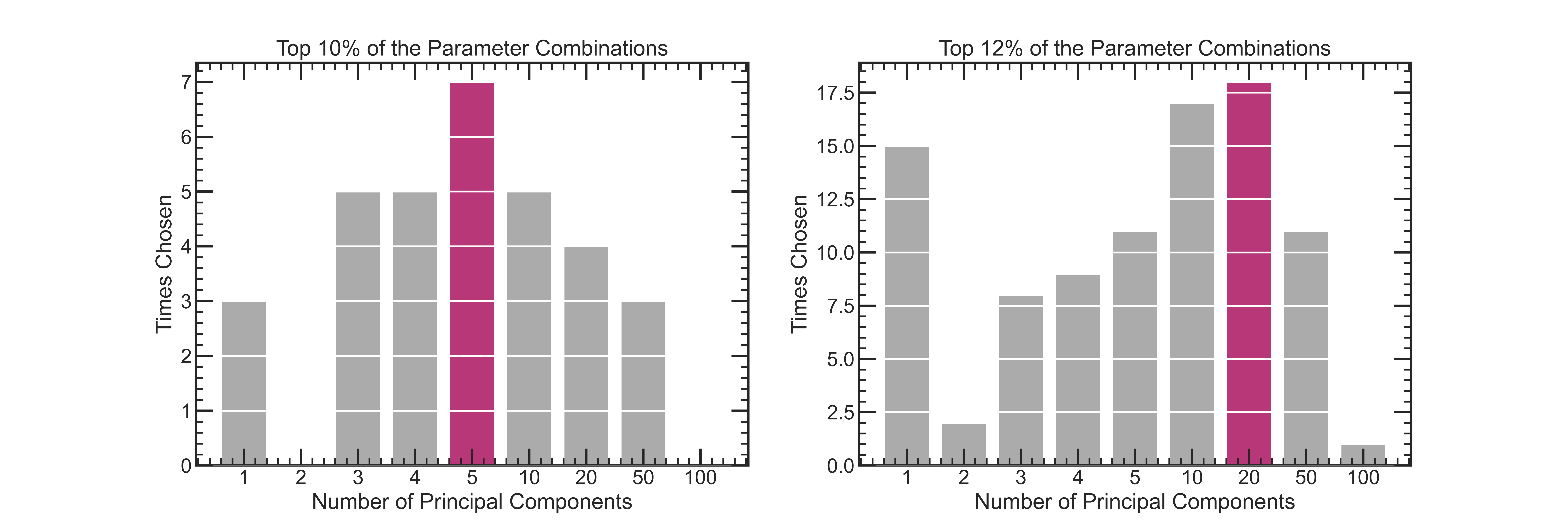}
    \caption{{Histograms showing the distribution of KL modes for all parameter combinations (choice of weights for image quality metrics and choice of KL modes for combination) where the four quantitative measures of similarity between parameter explorer heatmap structures for true and injected companions (described in detail in the text) were among the top 10\% (left) and top 12\% (right) of values according to all four measures. In the top 10\% of reductions, 5 KL modes was chosen most frequently. In the top 12\% of reductions, 20 KL modes was chosen most frequently.}}
    \label{fig:klhist}
    \end{center}
\end{figure*}

The metrics chosen varied, though again there was a preference for more than one image quality metric. We explored parameter combinations within the top 10\% and 12\% ``best" values for all four measures (the median, standard deviation, and sum of the difference maps, and the value of the difference map at the location of the Continuum injected planet peak), and found that in the top 10\%, 5 KL modes was chosen the most frequently, and in the top 12\%, 20 KL modes was chosen then most frequently (Figure \ref{fig:klhist}). Among all metric combinations that fell in the top 12\%, in 25 out of 26, at least 1 KL mode 5 or lower, and 1 KL model 10 or higher were selected. Therefore, we opted to use both 5 and 20. 

Informed by this analysis, we opted for a final ``best" collapse method of 5 and 20 KL modes and an equal weighting of all six image quality metrics.


We show the Continuum injected planet and true H$\alpha$ companion aggregate maps for this combination, as well as the difference map and a histogram of its values for this combination of image quality metrics and KL modes for all datasets in Figure \ref{fig:TFplanets}.

\begin{figure*}
    \centering
    \includegraphics[width=0.9\textwidth]{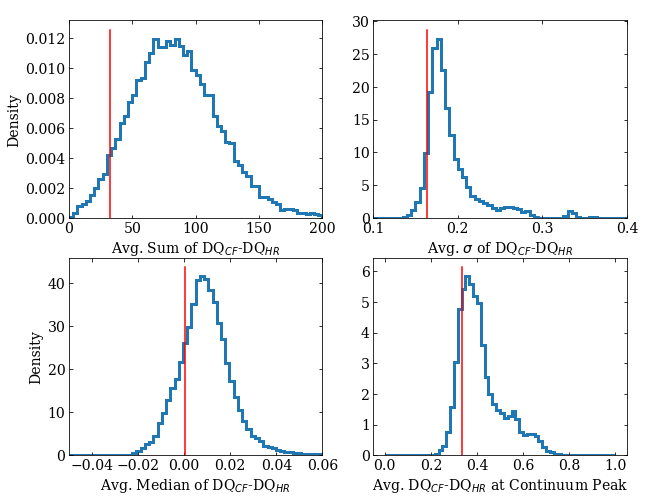}
    \caption{Histograms of summary statistics for Continuum injected planet - real H$\alpha$ companion difference maps. These maps are generated for each of 32,768 possible combinations of image quality metric and KL mode from the pyKLIP-PE output. For each map, the sum of the difference map (upper left), standard deviation of the difference map (upper right), median of the difference map (lower left) and the difference in the aggregate data quality (ADQ) metric score between the injected and true companions at the location of the peak (lower right) is computed. These quantities are averaged across the six HD142527 B detections and these averaged quantities are depicted in the histograms. The red vertical lines indicate the value of these statistics for our choice of ``best" collapse method.}
    \label{fig:metrichist}
\end{figure*}

The overall distributions of the standard deviation, sum, median, and difference at the peak of the difference maps for all collapse scenarios with the final choice marked in red is shown in Figure \ref{fig:metrichist}. 

\subsection{Computing Efficiency}

While a grid search of KLIP parameters applied to injected planets can improve detection quality, we recognize that it may not always be the most time-efficient choice. Note that the \texttt{pyKLIP-PE} algorithm takes hours to weeks to run on a single GAPlanetS dataset on an 8 core, 32 GB machine. Therefore, next steps in this exploration process will include investigating more systematically whether there are regions of KLIP parameter space that yield consistently poor detections and can be discarded. This would reduce processing time by decreasing the number of values tested. We caution that these parameters may be peculiar to the nature of the wavelength regime and instruments used for a given dataset.

\subsection{Future Directions}

Some parameters that were not explored in detail in the scope of this research may also be vital in optimizing detections. One example of this is the `highpass' \texttt{pyKLIP} parameter. Preliminary explorations of the highpass parameter show that it can change SNR by a factor of 3 or more. We find that using highpass values close to that of the PSF's FWHM usually result in high-quality detections. However, this relationship should be further explored, and perhaps even incorporated into the parameter exploration grid. 

These optimization techniques should also be tested on data from other telescopes. Our specific grid optimization technique is certainly biased towards GAPlanetS data, so it is vital to assess to what extent our conclusions are instrument or wavelength dependent.

In order to gain a full understanding of the detectability of accreting protoplanets embedded in disks, future analyses should also consider forward modeling of systems with a disk and planet combination.



\section{Summary and Conclusion} \label{conclusion}

In this paper, we demonstrate that a systematic approach to optimization of input parameters to the PSF-subtraction algorithm \texttt{pyKLIP} results in equivalent or higher SNR detections of companions in the Giant Accreting Protoplanet Survey (GAPlanetS) sample relative to using a single generic set of parameters across the survey dataset. We begin with six datasets of the HD 142527 system taken over the course of 5 years. We gauge the quality of our parameter selection method, which relies on {optimization of planetary signals injected into the Continuum images}, based on its ability to recover these known companions in as many datasets and at as high a SNR as possible.  

We introduce a grid search tool to optimize \texttt{pyKLIP} parameters using a number of post-processed image quality metrics. More specifically, we explore the role of the \texttt{pyKLIP} user-input parameters \texttt{movement}, \texttt{annuli}, and \texttt{numbasis} (KL modes) on the quality of post-processed images. To gauge image quality, we combine a number of metrics computed from the final signal-to-noise maps of the post-processed images. These metrics are: the peak (single pixel) SNR of the recovered planet, the average SNR of the positive pixels under the planet mask (r$\sim$0.5FWHM), the star/planet contrast achieved at the planet location, the number of false positive pixels between the inner working angle and control radius of the SNR map, and the quality of nearby combinations of PyKLIP parameters. 

The process of optimization utilized in this work is summarized as follows:

\begin{enumerate}
    \item Inject 2-8 synthetic planets into the \textit{Continuum} wavelength images for each GAPlanetS target in between the IWA and control radius 
    
    \item Fix the KLIP parameters \texttt{highpass} and \texttt{IWA} at values that were found to be universally reasonable (1 FWHM in most cases).
    
    \item Complete a coarse grid search of the \texttt{movement}, \texttt{annuli} and \texttt{numbasis} parameters with the \texttt{pyKLIP-PE} algorithm
    
    \item Compute SNR maps for each post-processed image and record various metrics for the ``quality" of each injected planet detection, namely: Peak SNR, Peak SNR neighbor quality, average SNR, average SNR neighbor quality, contrast, and number of $>$5$\sigma$ false positive pixels inside of the control radius. 
    
    \item Combine all six metrics and average among the 5 and 20KL mode reductions to arrive at a ``best" choice of KLIP parameters for a given set of Continuum injected planets
    
    \item Apply the best injected planet parameters to the $H\alpha$ data and record the Peak SNR of the real companion detections. If the injected planet was able to effectively mimic a real signal, then its optimal parameters should be well-approximated for a real planet in the same dataset.  
\end{enumerate}

We demonstrate that, relative to reductions with a generic fixed choice of KLIP parameter values, this simple grid search technique is able to reveal the HD 142527 B companion in every epoch and improve the SNRs of the detections by up to 1.2$\sigma$. 



This simple parameter grid search can help shape our understanding of how to find planets in an uninformed planet search. By showing that synthetic planets injected into images at a wavelength where true sources are expected to be dim can be used as a reasonable proxy for true planets at neighboring wavelengths, we can start to conceive of parameter optimization via more advanced mechanisms (such as neural networks). Furthermore, we have established a reliable and systematic method to select KLIP parameters without relying on the reality of the planetary signal itself.

This paper provides one solution to minimizing false positive protoplanet detections by developing a robust data-driven method for KLIP parameter optimization that does not rely on the reality of an apparent planetary signal. It introduces a new tool that will help high-contrast imaging surveys make the most of available data, and may even help reveal planets missed in previous explorations. 

\section{Acknowledgements}

\par KBF, WOB, and JA acknowledge funding from NSF-AST-2009816. LMC's work was supported by NASA Exoplanets Research Program (XRP) grants 80NSSC18K0441 and 80NSSC21K0397. WOB thanks the LSSTC Data Science Fellowship Program, which is funded by LSSTC, NSF Cybertraining Grant \#1829740, the Brinson Foundation, and the Moore Foundation; their participation in the program has benefited this work. KMM's work has been supported by the NASA XRP by cooperative agreement NNX16AD44G.

\par This paper includes data gathered with the 6.5 meter Magellan Telescopes located at Las Campanas Observatory, Chile.

\par JA would like to thank Sarah Blunt for her help reviewing the extended thesis version of this work, and her continued support and mentorship. JA especially thanks Sam and Captain Levi for their encouragement writing this paper, and their help bringing it to completion. JA also thanks Prof. Thayne Currie for his thoughtful and constructive feedback on this paper.

\par We thank the anonymous reviewer for their considerate and instructive review that greatly improved the quality of this paper.


\bibliography{Bibliography}{}
\bibliographystyle{aasjournal}

\end{CJK*}
\end{document}